\documentclass[aps,prb,twocolumn,superscriptaddress,floatfix,showpacs]{revtex4}

\usepackage{amssymb}
\usepackage{amsmath}
\usepackage{graphicx}
\usepackage{natbib}

\begin{document}

\title{Transport in Graphene: Ballistic or Diffusive?}

\author{Mario F. Borunda}
\email[Electronic address:\;]{mario.borunda@okstate.edu}
\affiliation{Department of Physics, Oklahoma State University, Stillwater, OK 74078, USA}
\affiliation{Department of Physics, Harvard University, Cambridge, MA 02138, USA.}
\author{H. Hennig}
\affiliation{Department of Physics, Harvard University, Cambridge, MA 02138, USA.}
\author{Eric J. Heller}
\affiliation{Department of Physics, Harvard University, Cambridge, MA 02138, USA.}
\affiliation{Department of Chemistry and Chemical Biology, Harvard University, Cambridge, MA 02138, USA.}

\date{\today}

\begin{abstract}
We investigate the transport of electrons in disordered and pristine graphene devices. 
Fano shot noise, a standard metric to assess the mechanism for electronic transport in mesoscopic devices, has been shown to produce almost the same magnitude ($\approx 1/3$) in ballistic and diffusive graphene devices and is therefore of limited applicability.
We consider a two-terminal geometry where the graphene flake is contacted by narrow metallic leads.
We propose that the dependence of the conductance on the position of one of the leads, a conductance profile, can give us insight into the charge flow, which can in turn be used to analyze the transport mechanism.
Moreover, we simulate scanning probe microscopy (SPM) measurements for the same devices, which can visualize the flow of charge inside the device, thus complementing the transport calculations.
From our simulations, we find that both the conductance profile and SPM measurements are excellent tools to assess the transport mechanism differentiating ballistic and diffusive graphene systems.
\end{abstract}


\maketitle
\section{Introduction}
As predicted by early band theory studies,\cite{Wallace_1947} charge carriers in graphene are chiral quasiparticles that have a linear Dirac-like dispersion relation resulting in fascinating electronic~\cite{CastroNeto_2009} and transport~\cite{DasSarma_2011} properties.
As a result of this linear energy-momentum relationship, the valence and conduction bands intersect at the Dirac points located at the Brillouin zone corners.
For neutral graphene, the chemical potential lies exactly at the Dirac points.
Use of a gate potential can shift the chemical potential away from the Dirac point and thus tune the charge density of the graphene device over a large range (as compared to other electronic materials). 
The high degree of control over the charge density and the linear dispersion relation are features that distinguish graphene from the two-dimensional electron gas in semiconductor heterostructures and other metallic conductors.
Understanding the mechanism for charge carrier transport in graphene in the low carrier density regime (the linear dispersion regime) and in the presence of disorder has received considerable attention.\cite{CastroNeto_2009,DasSarma_2011,Geim_2009,Rozhkov_2011}

The wide variety of fabrication techniques~\cite{Novoselov_2004, Berger_2006, Meyer_2007, Bolotin_2008, Du_2008, Dean_2010} has resulted in graphene devices that operate in different disorder regimes. 
The strength of the disorder charge carriers encounter can be quantified via the electron mobility.
The standard fabrication method of mechanical exfoliation and deposition of graphene flakes on SiO$_2$ substrates~\cite{Novoselov_2004} can result in good quality devices with mobilities approaching a few tens of thousands $cm^2 V^{-1} s^{-1}$, values comparable to those found in conventional semiconductor devices. 
Yet, the substrate provides a source of impurities~\cite{Adam_2007} and corrugations~\cite{Katsnelson_2007, Gibertini_2010} for the electrons to scatter, limiting the mean free path $l$ to about a hundred nanometers.
Etching away the substrate is routinely done to produce suspended graphene devices, removing some sources of scattering.
Etching followed by current annealing increases the carrier mobility in suspended samples to values exceeding 200,000 $cm^2 V^{-1} s^{-1}$, resulting in mean free paths comparable to the size of the device.\cite{Bolotin_2008, Du_2008}
Similar sample quality (80,000 $cm^2 V^{-1} s^{-1}$ carrier mobility) is obtained in devices where the substrate is single-crystal hexagonal boron nitride.\cite{Dean_2010}

It is apparent that replacing oxide based substrates by high-quality crystal substrates or suspended graphene results in devices with improved mobility and reduced electron scattering events.\cite{Bolotin_2008, Du_2008, Dean_2010}
For transport near or at the Dirac point, where the density of states vanishes, the conductivity has a minimum value ($4e^2/\pi h$), even for ideal pristine graphene.\cite{Tworzydlo_2006}
Experimentally, the phase coherent transport in graphene devices with short-and-wide geometries has been characterized as ballistic~\cite{Miao_2007, DiCarlo_2008, Danneau_2008} indicating evanescent wave transport.  
In an ideal (no impurities or defects) graphene device, the mean free path is longer than the size of the device resulting in ballistic conductivity. 
Thus, transport in graphene qualitatively changes from {\em metallic diffusive} in SiO$_2$ substrates to what has been characterized as {\em ballistic} in suspended systems and when using crystalline boron nitride substrates. 

Our paper is organized as follows. 
In Sec.~II we discuss the usual transport measures, emphasizing the null results obtained from the Fano shot noise, and present two alternative methods that can also estimate the transport mechanism. 
In Sec.~III we present the computational method used to model transport in two-terminal graphene devices.
In Sec.~IV we report results of extracting the transport mechanisms from modeled transport measurements.
We present a summary and conclusion in Sec.~V.

\section{Transport measures}
The possibility of several transport mechanisms makes it essential to develop measures that clearly distinguish and classify the different transport regimes.
In principle, Fano shot noise is the metric of choice to assess the mechanism for electronic transport in mesoscopic devices.
However, the Fano shot noise is of limited applicability, with the same value for the Fano factor found in disordered or pristine devices. 

\subsection{Shot noise and Fano factor}
Shot noise measurements can determine the statistics relevant for transport in mesoscopic conductors.~\cite{Blanter_2000}
Shot noise is a consequence of charge quantization and can be assessed by the Fano factor defined as the ratio of noise power and mean current, 
\begin{equation}
\mathcal{F}= \frac{\sum_{n=1}^N T_n(1-T_n)}{\sum_{n=1}^N T_n}. 
\label{eq:fano}
\end{equation}
From the above definition, the Fano factor should be zero for conventional ballistic systems, {\em i.e.} if perfect transmission is present.
It has also been established that for Poisson processes $\mathcal{F}= 1$ and for diffusive metallic conductors $\mathcal{F}=1/3$.\cite{Beenakker_1992}

However, the Fano factor calculated for pristine graphene,\cite{Tworzydlo_2006,Schomerus_2007,Barraza_2012} disordered systems,\cite{Louis_2007,San-Jose_2007} samples with substrate roughness,\cite{Lewenkopf_2008} and for transport along an n-p junction~\cite{Cheianov_2006} all have similar magnitude ($\approx 1/3$) to that found in diffusive metallic conductors, making the Fano shot noise a cumbersome measure of the transport mechanism in graphene devices.\cite{DasSarma_2011}
Calculations in pristine graphene predict that the Fano factor at the Dirac point is identically 1/3 when 
(1) a wide device (where the aspect ratio W/L is above 3) is contacted by doped graphitic leads,~\cite{Tworzydlo_2006} 
(2) when the leads are quantum wires creating an effective-contact model simulating the metallic lead/graphene interface,~\cite{Schomerus_2007} and 
(3) for graphene junctions with realistic metal contacts.~\cite{Barraza_2012}
Away from the Dirac point, the Fano factor also indicates diffusive metallic transport.~\cite{Schomerus_2007,Barraza_2012}
A slightly smaller yet universal value ($\mathcal{F}= 0.295$) was found numerically for disordered systems.~\cite{Louis_2007, San-Jose_2007}
When disorder originates mainly from roughness in the substrate, the Fano factor lies slightly above the 1/3 value and has been shown numerically to increase slightly with disorder.~\cite{Lewenkopf_2008}
Transport along an n-p (electron rich/hole rich) junction is selective of those quasiparticles that approach the n-p interface almost perpendicularly and results in a Fano factor of $\mathcal{F}= 1-\sqrt{1/2}$.~\cite{Cheianov_2006}
Again, in all of these different situations, the Fano factor measurement results in 1/3 or a value numerically close to this number.

Experiments have found similar results. 
The Fano factor measured in clean devices~\cite{Danneau_2008} is close to the value of 1/3 found analytically for ballistic samples.~\cite{Tworzydlo_2006}
Measurements in disordered devices~\cite{DiCarlo_2008} follow the numerical trends of Ref.~\onlinecite{Lewenkopf_2008}.
These similar shot noise values warrant developing an alternative quantitative understanding of the carrier dynamics crucial for testing the transport regime of graphene devices.

\subsection{Conductance profiles}

Our quest for a better measure to assess transport mechanisms begins with a simple question: How does the conductance change with respect to the vertical displacement of one of the leads?
By calculating the conductance between two metallic contacts, we model transport in two-terminal devices. 
We extract the conductance profile from the dependence of the conductance on the position of one of the leads, and use this to distinguish the mechanism of transport.
In particular, we study the transport properties of clean and disordered graphene devices contacted by narrow metallic leads where one of the two leads can be moved along the edge of the device. 
Moreover, we simulate scanning probe microscopy (SPM)~\cite{Topinka_2000} measurements for the same devices, which sheds light on the charge flow \emph{inside} the device.~\cite{Topinka_2001}

Diffusive transport in disordered systems is based on the charge carriers scattering multiple times off impurities or boundaries as they traverse the system.
This mechanism can be described classically by a random walk.
In ballistic transport, the charge carriers traverse the system with minimal scattering. 
Ballistic transport is expected in an ideal graphene strip, given that the crystal lattice has no defects and no impurities are present.
Yet for ideal graphene, the dynamics of the electrons produce the same shot noise as that found in classical diffusion.~\cite{Tworzydlo_2006}
Determining the transport mechanism in graphene devices is important given that transport experiments are possible in the quasi-ballistic limit, that is, where the mean free path is of the order of the size of the system.

Here, we argue that the transition from diffusive to ballistic transport in graphene, along with the limiting cases, can be quantified by alternative methods based on measuring the flow of charge in the system. 
First, the conductance profile, that is, the conductance as a function of the vertical displacement of one of the leads, can give a measure of the flow inside the device.
For the case of ballistic transport, scattering does not impede the flow of charge carriers and the conductance profiles are well described by a linear (triangular) fit due to the convolution of the square windows created by the two leads.
In the case of diffusive transport, the charge carriers experience several scattering events and the transport is then well described by Brownian motion with a drift.
Thus, charge flow will be a Gaussian function characteristic of diffusion; for this reason, the conductance profile can be fitted to a Gaussian function.

\section{Computational method}

\begin{figure}
\begin{center}
\includegraphics[scale=0.14]{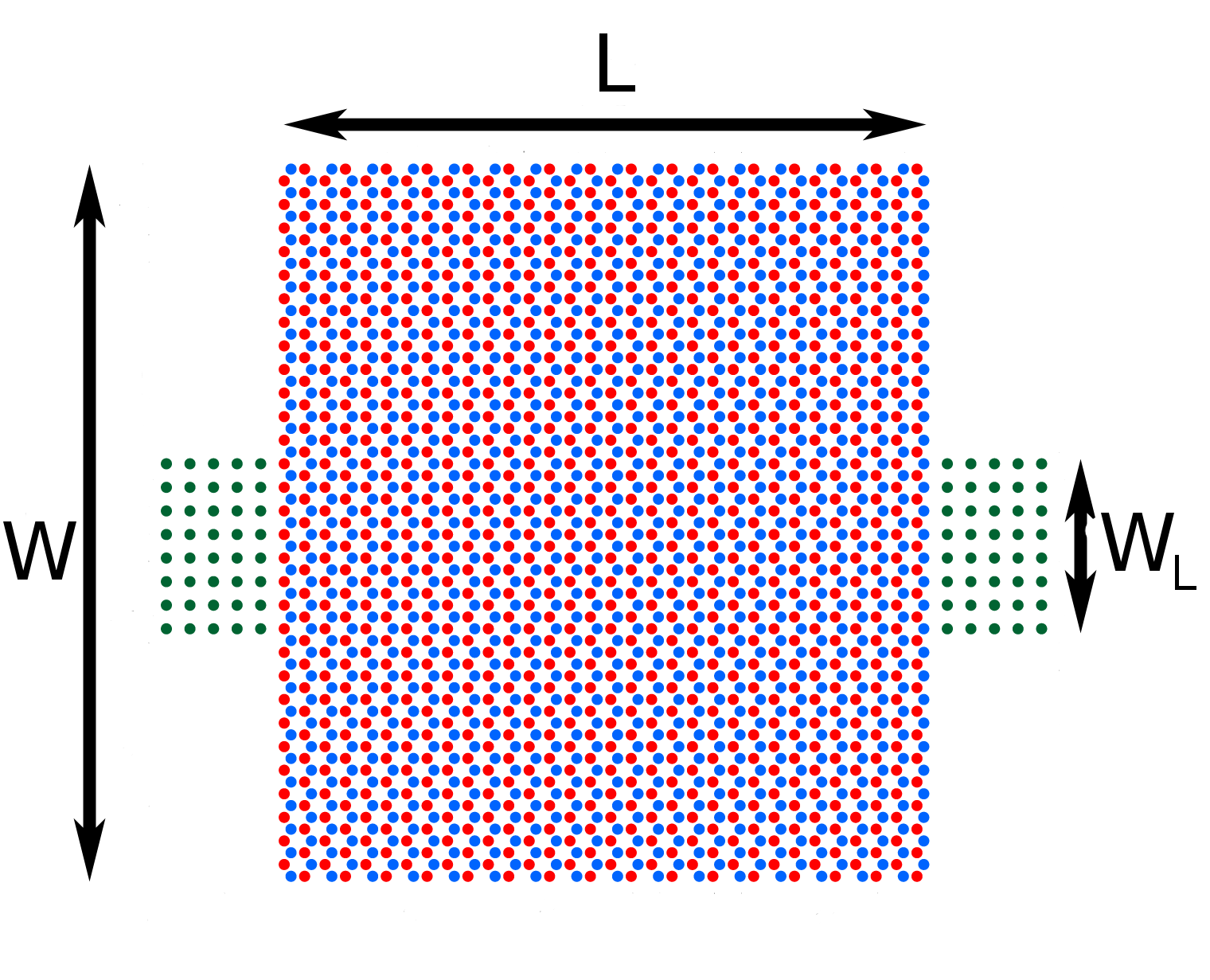}
\caption{\label{fig1} (Color online) Schematic setup for the transport calculation. 
A rectangular graphene flake, depicted by red and blue atomic sites arranged in a honeycomb lattice, is contacted between two narrow metallic contacts, depicted by the green atomic sites arranged in a square lattice, forming a normal-conductor/graphene/normal-conductor junction. 
This schematic is not to scale; simulated devices were considerably bigger with $\sim700$ carbon atoms along the zigzag edge.}
\end{center}	
\vspace{-0.50 cm}
\end{figure}

We use the nonequilibrium Green's function (NEGF) formalism~\cite{Haug_2008} to calculate the transport properties of a normal-conductor/graphene/normal-conductor junction, given that in experiments the electronic contacts are usually metallic. 
The metallic lead/graphene interface we have investigated follows that of Robinson and Schomerus.~\cite{Robinson_2007}
A schematic of the geometry of the devices investigated is shown in Fig.~\ref{fig1}, where rectangular (width W and length L) graphene flakes have armchair boundaries along their longitudinal direction and narrow metallic leads connected along the zigzag boundary. 
The two metallic terminals of width W$_{\mbox{L}}$ are modeled by semi-infinite square lattice regions that have a quadratic dispersion relation. 

The tight-binding model Hamiltonian of the device is given by
\begin{equation}
H = \sum_i \epsilon_i  c^\dagger_i c_i - \sum_{\langle i,j \rangle} \gamma_{ij} c^\dagger_i c_j,
\end{equation}
where  $\epsilon_i$ is the on-site potential of the lattice site $i$, $c_i (c^\dagger_i)$ is the annihilation (creation) operator acting on site $i$, the second sum is over nearest neighbors $\langle i,j \rangle$, and $\gamma_{ij}$ denotes the hopping matrix elements. 
The on-site potential changes due to contributions from impurities and the gate voltage applied to the device, $\epsilon_i = \epsilon_{imp} + \epsilon_{gate}$. 
The graphene section consists of atomic sites placed in the hexagonal lattice with lattice constant $a$.
As outlined in Ref.~\onlinecite{Robinson_2007}, metallic contacts are modeled as a region with a square lattice arrangement with lattice constant $a_L = \sqrt{3} a$, matching the A(B)-atom in the zigzag interface at the left(right) of the graphene flake.

Disorder is introduced in two different ways. 
To generate edge-disordered samples, we randomly remove 30\% of carbon atoms from sites located in the three outer atomic layers of the device.~\cite{Louis_2007} 
This type of disordered edge without disorder puddles in the bulk of the system is a reasonable model for suspended graphene devices.

Bulk disordered in graphene devices is linked to the presence of charged impurities in the substrate.~\cite{Rycerz_2007} 
To generate such disorder potentials, $N_{imp}$ lattice sites are selected at random from the total number of atomic sites ($N_{tot}$) in the device.
The position $R_i$ of each of the $N_{imp}$ lattice sites has an on-site potential amplitude $V_i$ chosen randomly from the interval $(-\delta, \delta)$ and smoothed out over a range $\xi$ by convolution with a Gaussian,

\begin{equation}
\epsilon_{imp}(R_i) = \sum_{i=1}^{N_{imp}} V_i \exp \left( - \frac{ \left| r - R_i \right|^2}{2 \xi^2} \right). 
\end{equation}
The range of the convolution is important given that in the extreme case of $\xi << a$ the atomic scale disorder potential would break the A-B symmetry caused by having two atoms in the unit cell. 
For our calculations we assume that $\xi = 2 a$ resulting in a short-range potential (when compared to other length scales in the system) that varies smoothly on the atomic scale and suppresses the effect of intervalley scattering.
The parameters $N_{imp}$ and $\delta$ determine the mean free path $l$ of the disordered system.
Using the Born approximation, the mean free path can be quantified as~\cite{Rycerz_2007, Suzuura_2002}
\begin{equation}
\label{eq:mean-free-path}
l = \frac{4}{k_F K_0},
\end{equation}
where $k_F$ is the Fermi wave vector and $K_0$ is the dimensionless correlator given by
\begin{equation}
K_0 = \frac{L W}{(\hbar v N_{tot})^2} \sum_{i=1}^{N_{tot}} \sum_{j=1}^{N_{tot}} \langle \epsilon_{imp}(R_i) \epsilon_{imp}(R_j) \rangle.
\end{equation}
For the type of bulk disordered systems considered here ($\xi << L, W$), Rycersz {\em et al.}~\cite{Rycerz_2007} have found that
\begin{equation}
K_0 = \frac{\sqrt{3}}{9} \frac{\delta^2}{\gamma^2} \frac{N_{imp}}{N_{tot}} \kappa,
\end{equation}
where
\begin{equation}
\kappa = \frac{1}{N_{imp}} \sum_{i=1}^{N_{imp}} \sum_{j=1}^{N_{tot}} \exp \left( - \frac{ \left| r - R_i \right|^2}{2 \xi^2} \right).
\end{equation}

The NEGF formalism~\cite{Haug_2008,Nicolic_2010} is a sophisticated framework for obtaining the transmission and other quantities in realistic devices. 
The retarded Green's function in the atomic orbital basis is given by
\begin{equation}
G^r(E) = \left[ E - H - \Sigma(E) \right]^{-1},
\end{equation}
where the non-Hermitian self-energy matrix 
\begin{equation}
\Sigma(E) = \Sigma_L(E) + \Sigma_R(E),
\end{equation}
introduces the effect of attaching leads to the left and right ends of the device. 
The energy $E$ tunes the Fermi level from the Dirac point ($E=0$) to any charge density induced by the gate voltage in the device.
The self-energies determine the escape rates of electrons from the device into the semi-infinite ideal leads.
Using the Landauer formula~\cite{Landauer_1957}, it is possible to obtain the transmission function
\begin{equation}
T(E, V_{ds}) = Tr \left[ \Gamma_R(E,V_{ds}) G^r \Gamma_L(E,V_{ds}) G^a \right],
\end{equation}
where $ G^a(E) = (G^r(E))^\dagger $ is the advanced Green's function.
The matrices 
\begin{equation}
 \Gamma_{p}(E,V_{ds}) = i \left[ \Sigma_{p}\left(E-\frac{e V_{ds}}{2}\right) - \Sigma^\dagger_{p}\left(E+\frac{e V_{ds}}{2}\right) \right] 
\end{equation}
introduce a level broadening due to the coupling of the leads and a source-to-drain voltage given by $V_{ds}$.   
The Fano factor [Eq.~(\ref{eq:fano})] for a particular energy and $V_{ds}$ is calculated using the following expression,
\begin{equation}\label{eq:fano2}
F = 1- \frac{Tr \left[ \Gamma_R G^r \Gamma_L G^a \Gamma_R G^r \Gamma_L G^a \right]}{T}.
\end{equation}
Finally, the current in two-terminal devices can be obtained from the Landauer formula
\begin{equation}
I(V_{sd}) = \frac{2e}{h} \int_{-\infty}^\infty dE~T(E) \left[ f(E-\mu_L) - f(E-\mu_R) \right]
\end{equation}
where the energy window is defined from the difference of the Fermi functions of the macroscopic reservoirs where the leads terminate. 
In our calculations we assume the linear response limit ($V_{ds} \rightarrow 0$), where the relationship between conductance $G$ and current is given by $I = G V_{ds}$.
\section{Results and Discussion}
\subsection{Fano shot noise}
We investigate rectangular graphene devices with a two-terminal geometry where the source and drain leads are assumed to be perfect ballistic and metallic conductors. 
This geometry has been investigated previously in the context of quantum scars in graphene~\cite{Huang_2009} and for the geometry-dependent conductance fluctuations in graphene quantum dots.~\cite{Huang_2011}
For all results presented here and as illustrated in Fig.~\ref{fig1}, the edges of the graphene along the transport direction are in the armchair configuration.

\begin{figure}[t]
\begin{center}
\includegraphics[width=0.55\textwidth]{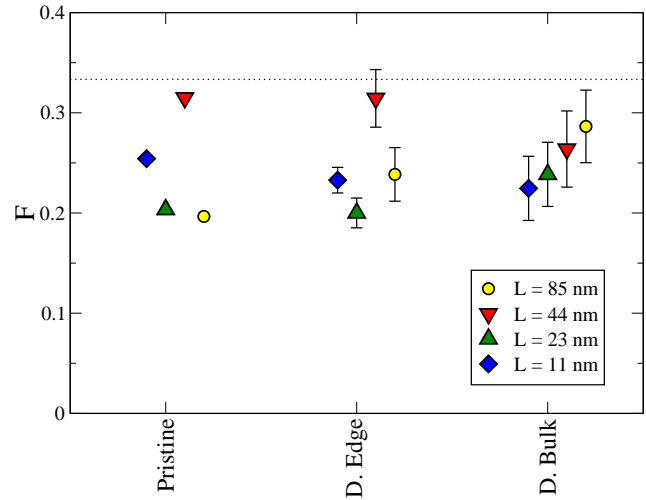}
\caption{\label{fig:fano} (Color online) Fano factor at a Fermi energy $E_F=0.5 \gamma$ (with hopping matrix element $\gamma$).
The limit $\mathcal{F} = 1/3$ as $W/L \rightarrow \infty$ is shown as a dotted line. 
The data points of the disordered systems have error bars showing the variance of the value over several realizations. 
Each of the data points corresponds to different system lengths (L = 11, 23, 44, and 85 nm) while the widths of the graphene device and of the leads are kept constant at W = 74 nm and W$_{\mbox{L}}$ = 10 nm, respectively.
The mean free path is $l = 27$ nm in the system with bulk disorder.}
\end{center}	
\vspace{-0.50 cm}
\end{figure}

We use Eq.~\ref{eq:fano2} to calculate the Fano factor for pristine, edge disordered, and bulk disordered graphene devices.
As presented in Fig.~\ref{fig:fano}, our results are similar to those found in the literature.\cite{Tworzydlo_2006,Schomerus_2007,Barraza_2012,Louis_2007, San-Jose_2007,Lewenkopf_2008}
The Fano factors found are near the theoretical value $\mathcal{F} = 1/3$ that applies to pristine graphene as $W/L \rightarrow \infty$ near the Dirac point limit.
As Towrzydlo {\em et. al.} reported,\cite{Tworzydlo_2006} the $\mathcal{F} = 1/3$ value is a theoretical maximum for armchair edge graphene devices, increasing the charge density reduces the value of the Fano factor.

For pristine devices, the Fano factor does show a trend as a function of the system size.
The longer device calculated (83 nm) presented the lowest value of $\mathcal{F}$ while the largest value corresponds to the 44 nm graphene sample.  
Likewise, the Fano factor for the graphene devices with disordered edges do not exhibit a trend in the sizes studied.
However, the fluctuations in $\mathcal{F}$ from the mean value, illustrated with the standard deviation (error bars), show that as the devices get longer, the deviation from the mean increases.
In contrast, for diffusive graphene, we find that the value of the Fano factor does show a monotonic dependence as a function of the $W/L$ ratio, approaching the $\mathcal{F} = 1/3$ value in the larger devices. 
Yet, given that the shot noise metric is not significantly different from the value found in classical diffusive transport, even in the narrow-leads geometry, the Fano factor is an inadequate metric for predicting the transport mechanism.

\subsection{Transport and the profile of the conductance}

\begin{figure}[t]
\begin{center}
\includegraphics[width=0.37\textwidth]{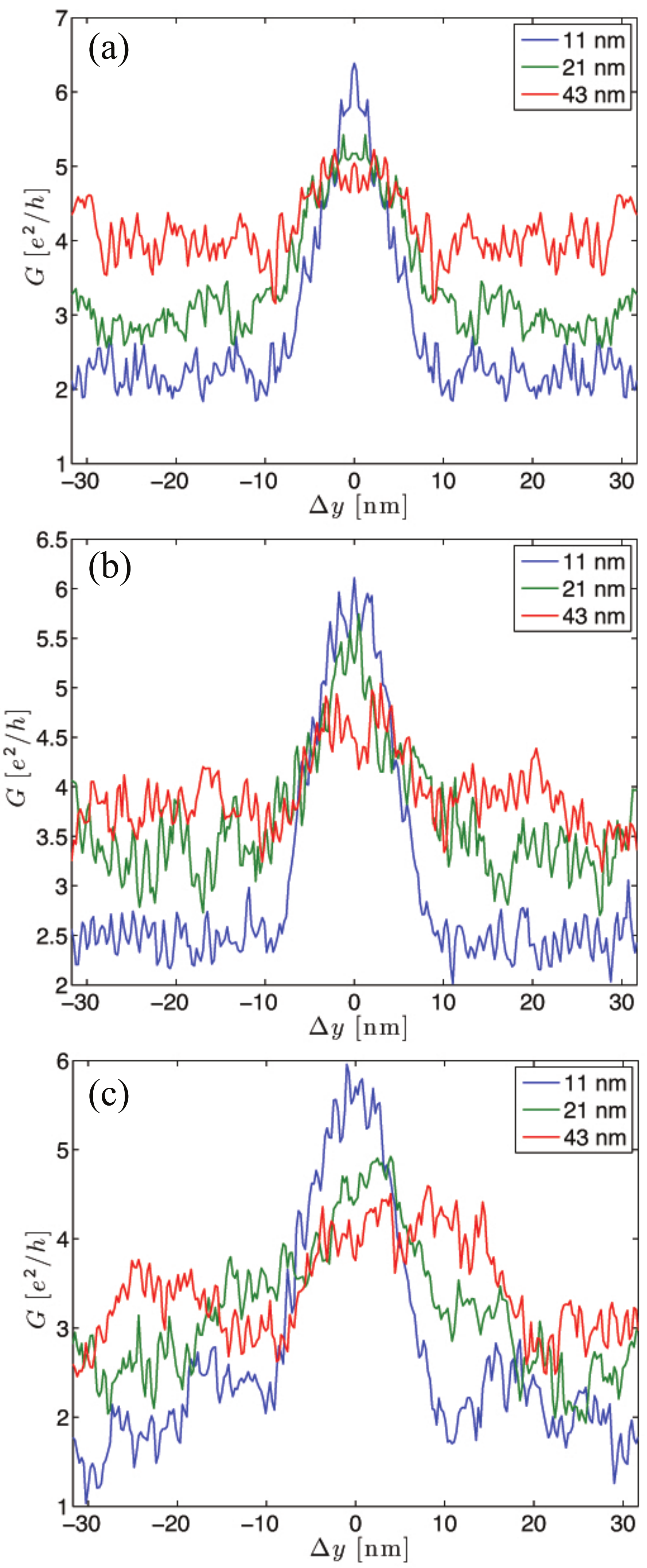}
\caption{\label{fig2} (Color online) Profile of the conductance for pristine, edge disordered, and bulk disordered systems. The conductance versus displacement of one of the leads from the center of the device  $G(\Delta y)$ is calculated at a Fermi energy $E_F=0.5 \gamma$ (with hopping matrix element $\gamma$).
Each of the curves corresponds to different system lengths (L = 11, 21, and 43 nm) while the width of the graphene device and of the leads are kept constant at W = 74 nm and W$_{\mbox{L}}$ = 10 nm, respectively. (a) Clean pristine system. (b) System with edge disorder. (c) System with bulk disorder where the mean free path is $l = 27$ nm.}
\end{center}	
\vspace{-0.50 cm}
\end{figure}

In order to visualize the charge flow at the edge of the sample, we calculate the profile of the conductance $G(\Delta y)$, that is, 
the conductance as a function of the position of the drain lead with respect to the source lead that remains fixed.
In Fig.~\ref{fig2}, we present the conductivity as a function of the displacement of the drain lead $\Delta y$ for several system lengths (L=11, 21, 43 nm) and for the three models considered here.
Both the width of the device and the width of the leads are kept constant at W = 74 nm and W$_{\mbox{L}}$ = 10 nm, respectively. 
Thus, the schematic shown in Fig.~\ref{fig1} is not representative of the size of the devices considered in this work as  our calculation are considerably larger with about 700 atomic sites along the zigzag edge (width of system).
Numerical calculations of these system sizes require the use of efficient recursive Green's function methods. 

While the displacement of one of the leads from the center of the device is a theoretical construct that can be easily investigated numerically, the experimental implementation is not trivial. 
Our proposal of ``moving the leads" could be realized by fabricating several samples where one of the leads is attached at a different location in each device.
Another possibility is to replace one of the metallic leads by the tip of a scanning tunneling microscope. 
When the tip is brought close enough to the edge, the tunneling current could be measured as a function of the position of the scanning tunneling microscope.

Figure~\ref{fig2}(a) corresponds to pristine graphene devices,
Fig.~\ref{fig2}(b) corresponds to realizations with edge disorder, and 
Fig.~\ref{fig2}(c) to systems where the presence of bulk disorder would form electron-rich and hole-rich puddles~\cite{Martin_2008} and the mean free path is $l = 27$ nm, as estimated from the Born approximation.
Figure \ref{fig2}(a) shows that the maximum conductance is associated with the two leads being collinear ($\Delta y \approx 0$).
For a short device the maximum conductance occurs at $\Delta y = 0$ and as the length of the system in increased there are local maxima near $\Delta y = 0$.
Increasing the length of the device reduces the conductance peak due to the increased number of reflections at the boundary of the sample.     
Yet, we find an envelope of maximum conductance when the vertical positions of the two leads overlap with each other.
Similar features are seen in the conductance profiles for the disordered samples.
We find that for all three systems the width of the central peak is approximately $2W_L$, corroborating the idea that the peak in the displacement conductance is due to the overlap of the two leads. 

Although our interest is to extract the transport mechanisms from the conductance profiles, there are certain features visible in the conductance curves of Fig.~\ref{fig2} that are worth explaining such as the fluctuations near the edges of the conductance curves. 
We carefully checked that these fluctuations are not only present for individual conductance curves and remain after averaging over an energy window. 
Thus, the fluctuations are consistent with universal conductance fluctuations (UCFs),~\cite{Lee_1985, Altshuler_1985} which we therefore review.

\begin{figure*}[t]
\begin{center}
\includegraphics[width=0.75\textwidth]{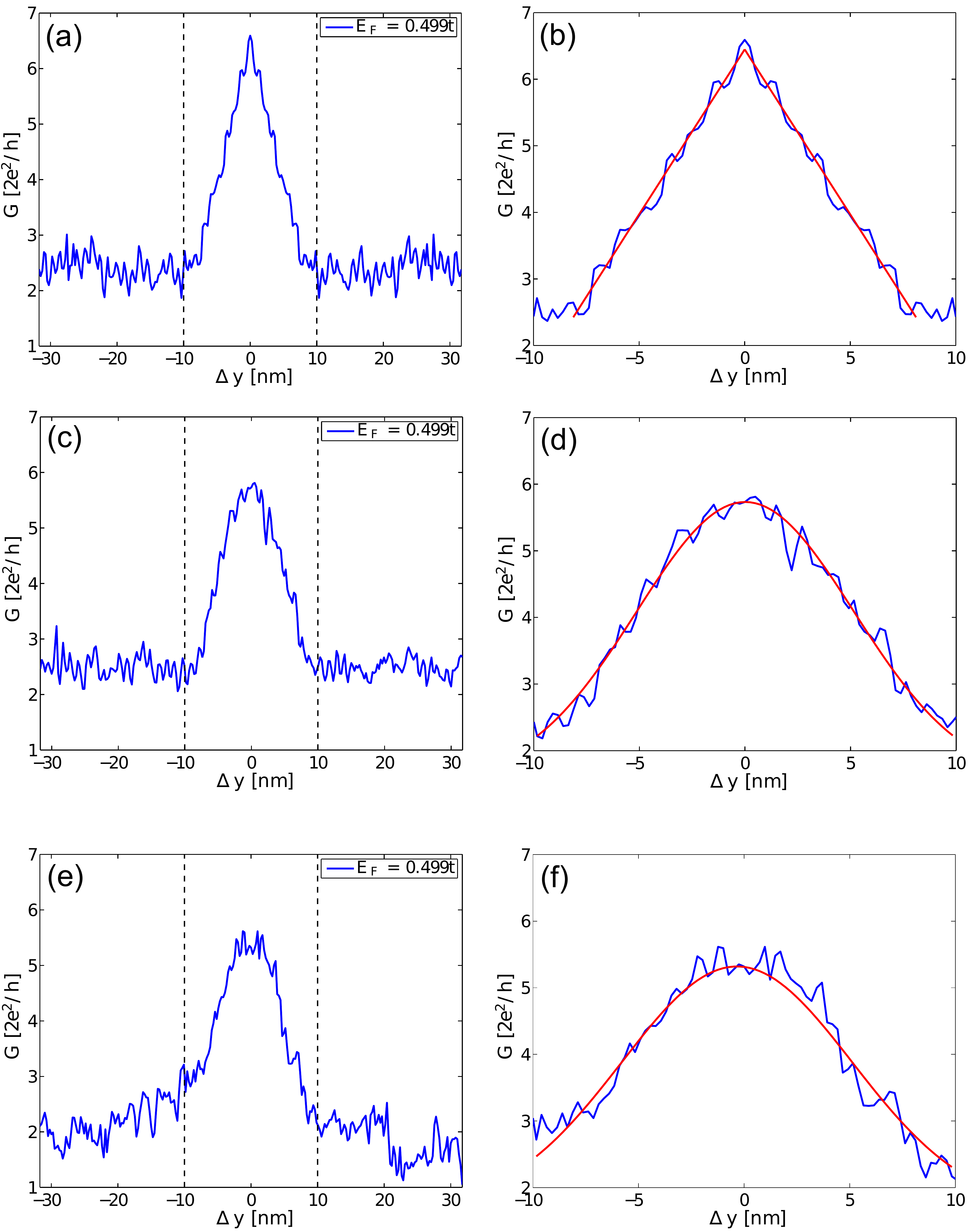}
\caption{\label{fig3} (Color online)
The transport mechanism is classified according to the shape of the central peak (of width $2W_L$ centered around $\Delta y$=0) of the conductance profile curves. 
$G(\Delta y)$ for the pristine system (a,b) is better fit by a triangular fit indicative of ballistic transport. 
In systems with edge disorder (c,d) and bulk disorder (e,f),  the central peak is well described by a Gaussian envelope pointing to diffusive transport.
The dashed lines indicate the region where the two leads have an overlap.
Parameters for all calculations are $E_F=0.499 \gamma$, $L=11\,$nm, W~=~74 nm and W$_{\mbox{L}}$ = 10 nm.}
\end{center}	
\end{figure*}

In a disordered mesoscopic conductor where the system is of comparable size or smaller than the phase coherence length of the charge carriers but large compared to the average impurity spacing, the transmission of carriers is affected by interference of many different paths through the system. 
As these paths are typically long compared to the wavelength of the charge carriers, the accumulated phase along the paths changes randomly in response to variation in an external parameter (e.g., the magnetic field or gate voltage). 
This results in a random interference pattern and reproducible fluctuations in the conductance of a universal magnitude on the order of $2e^2/h$.~\cite{Lee_1985, Altshuler_1985}
UCFs can also be created by the displacement of a single scatterer.~\cite{Feng_1986, Altshuler_1985b, Borunda_2011} 
But how can UCFs occur in a ballistic device? 
The role of disorder in providing a distribution of random phases can as well be taken by chaos. 
Thus, ballistic mesoscopic cavities like quantum dots in high mobility two-dimensional electron gases that form chaotic billiards show the same universal fluctuations.~\cite{Marcus:1992ug,Jalabert_90_PRL,Baranger:1993wl}
Remarkably, experiments in graphene quantum dots found strong indications of chaos in billiard systems.~\cite{Ponomarenko:2008ty}

In Fig.~\ref{fig3} we presents our conductance profile analysis for the 11 nm devices. 
A striking feature of these conductance profiles is the apparent fit of the peaks in the displacement conductance curves to either triangular or Gaussian functions.
The conductance profile for a pristine device of length $L=11\,$nm is shown in Fig.~\ref{fig3}(a) and the peak is presented in Fig.~\ref{fig3}(b).
The envelope of the peak in the conductance is well described by a triangular fit (red line). 
In the case of conventional ballistic transport, the classical expectation for the charge flow at the edge is a triangular shape of width $2W_L$. 
This triangular shape is the result of the convolution of two square windows, where the window sizes are given by the lead width $W_L$. 
Gaussian fits for the conductance profile (not shown) differ significantly particularly near the cusp of the curve.

Diffusive transport is based on the multiple scattering paths taken by charge carriers as they transverse a device.
In Fig.~\ref{fig3}(c) we present the conductance profile for an edge disordered system of length $L=11\,$nm and a close-up of the peak with a fit shown in Fig.~\ref{fig3}(d).
In this case, and in contrast with the wide and ballistic device, the data for the edge-disordered device is best fitted by a Gaussian curve, where the envelope of the conductance profile is the result of the spatial overlap of the leads.
Similarly, a Gaussian curve describes well the conductance profile of a diffusive system, as seen in Figs.~\ref{fig3}(e) and \ref{fig3}(f).
We note that Barthelemy {\em et al.}~\cite{Barthelemy_2008} used a similar approach to distinguish normal diffusion (reflected by a Gaussian profile) from the anomalous transport associated with L\'evy transport. 
\begin{figure*}
\begin{center}
\includegraphics[width=0.95\textwidth]{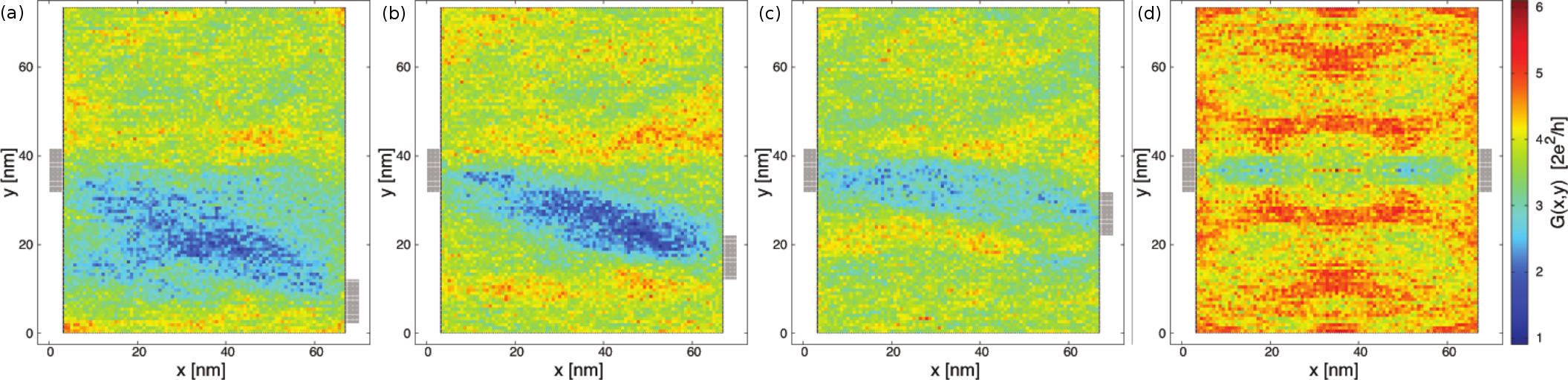}
\caption{\label{deltaY-scan} (Color online)
Charge flow in pristine graphene obtained from the SPM conductance map calculations. 
The maps of conductance as a function of SPM tip position, $G(x,y)$, show the charge flow in the graphene device.
Each snapshot presents the conductance map for different position of the right lead, starting from left at the bottom portion of the device, towards the right where the right lead is located at the same height of the left lead.
We can see that the majority of the flow is over a path connecting the left and right lead.
When the two leads are collinear, there are several bands of enhanced conductance due to the interference caused by paths that scatter from the tip and those that go from one lead straight to the other lead.
The Fermi energy is set to $E_F=0.5 \gamma$.
}
\end{center}	
\vspace{-0.5 cm}
\end{figure*}
While we have been able to account for the features of the envelope of the central peak of width $2W_L$ in a quasi-classical way, the conductance profile curves exhibit several quantum interference effects that need a different interpretation. 
Along with UCFs, Fabry-P\'erot conductance interferences show up as regularly spaced fluctuations seen in the conductance profiles of the pristine systems.

\subsection{Charge flow}

Scanning probe microscopy (SPM)~\cite{Topinka_2000} has recently been used to image mesoscopic transport effects such as universal conductance fluctuations~\cite{Berezovsky:2010wl} and weak localization~\cite{Berezovsky:2010wr} in graphene devices.
SPM has also been used to image current transport in a graphene quantum dot connected to leads through two small constrictions, imaging conductance resonances of the quantum dot and observing localized states.~\cite{Schnez_2010}
The tip of the cryogenic SPM capacitively couples to the graphene device inducing a movable scatterer. 
Conductance maps created via SPM have been proposed as a way to probe the chiral nature of the charge carriers in graphene.~\cite{Braun_2008}
The work carried out by Berezovsky {\em et al.}~\cite{Berezovsky:2010wl, Berezovsky:2010wr} and Schnez {\em et al.}~\cite{Schnez_2010} probed the regime of coherent diffusive transport in graphene.
In light of the above, it is interesting to consider the possibility of using SPM to gain insight into the transport mechanism by imaging the flow of charge carriers -- particularly, the ballistic to diffusive crossover regime in graphene devices.

We simulate the effect of the capacitively-coupled tip of the SPM on the graphene device as a point charge $q$ located a height $a$ above the substrate.
The charge induces a local charge density perturbation given by
\begin{equation}
 n(\rho) = \frac{-q a}{2 \pi (\rho^2 + a^2)^{3/2}}
\end{equation}
where $\rho$ is the in-plane radial coordinate away from the position of the point charge. 
The charge $q$ is chosen to yield an rms charge density $n \approx 4 \times 10^{11}~e~\mbox{cm}^{-2}$, in agreement with the observed charge puddles in graphene.  
The tip height $a$ controls the width of the induced charge density; we have chosen $a = 10$ nm, a routinely used tip-to-substrate distance in SPM experiments which results in a half-width for the induced density puddle of about 20 nm.
 
During SPM measurements, a transport measurement is carried out while the tip of the microscope scans the over the device.
The resulting conductance maps are generated by calculating the conductance as a function of SPM tip position $G(x,y)$ and rastering the tip position in a plane above the device. 
The effect of the SPM tip is included by adding to the on-site potential of each of the lattice sites the energy due to the charge density perturbation created by the tip.
The total on-site energy is obtained from the contributions of the perturbation from the tip, the effect from the gate voltage, and the impurities present, $\epsilon_i = \epsilon_{imp} + \epsilon_{gate} + \epsilon_{tip}$.    

In the previous section we extracted conductance profiles that allowed the classification of transport as ballistic or diffusive. 
To explain these observations and to obtain an intuitive picture of the transport mechanisms, we explore charge flow as a function of the position of one of the leads.
As in the conductance profile calculations, the injection lead remains fixed in the left edge of the sample.
The two terminals are identical. 
As seen in the panels of Fig.~\ref{deltaY-scan}, the SPM conductance maps exhibit a significant change in conductance when the charge flow between the leads is obstructed or enhanced by the presence of the perturbation induced by the scanning probe.

The conductance map shown in Fig.~\ref{deltaY-scan}(a) was obtained when the right lead is positioned near the bottom edge of the device.
As the scanning probe rasters along the top half of the device, the change in conductance is minimal as reflected by the mostly uniform conductance in that region of the image.
However, as the tip scans the bottom half of the device, the conductance map shows a region of lower conductance that connects the left lead to the right lead, as indicated by the large blue feature across the bottom of the image.
In this case, the trans-conductance is reduced when the local perturbation induced by the tip of the SPM is over a region of considerable charge flow.
In Fig.~\ref{deltaY-scan}(b), the right lead contacts the device at a slightly higher position.
Consequently, the blue feature corresponding to the region of charge flow is now at a higher position, ``following" the position of the right lead. 
Furthermore, there are two regions of higher conductance in the image, corresponding to an increase of the conductance due to the presence of the local perturbation.
The higher conductance (red) bands are a consequence of the redirection of the charge flow that without the tip being present would not have contributed to the conductance.
Thus, the tip redirects charge into the drain lead that otherwise would not have exited the device.

Similarly, placing the right lead closer to the central region of the device (Fig.~\ref{deltaY-scan}(c)), we find that the position of conductance features are correlated with the path starting at the left lead and ending in the right lead.
In all images, the effect of narrow lead constrictions is seen from the size of the band of near constant conductance.
Surprisingly, the feature does not become significantly wider as the charge flows away from the source lead.
All of the panels present conductance fluctuations of order $\sim \pm 2 e^2/h$.

The conductance map presented in Fig.~\ref{deltaY-scan}(d) corresponds to a device where the leads are collinear ($\Delta y = 0$). 
Here, the conductance is higher than in the ones to the left.
Along with the central band of conductance drop, Fig.~\ref{deltaY-scan}(d) also shows regions of constructive interference along bands parallel to the transport.  
The limit to the size of the features present in the conductance maps is proportional to the Fermi wavelength of the system~\cite{Topinka_2000,Topinka_2001}, which for graphene is inversely proportional to the Fermi energy given by
\begin{equation}
 \lambda_F = 2 \pi \frac{v_F}{E_F}.
\end{equation}

\begin{figure}
\begin{center}
\includegraphics[width=0.45\textwidth]{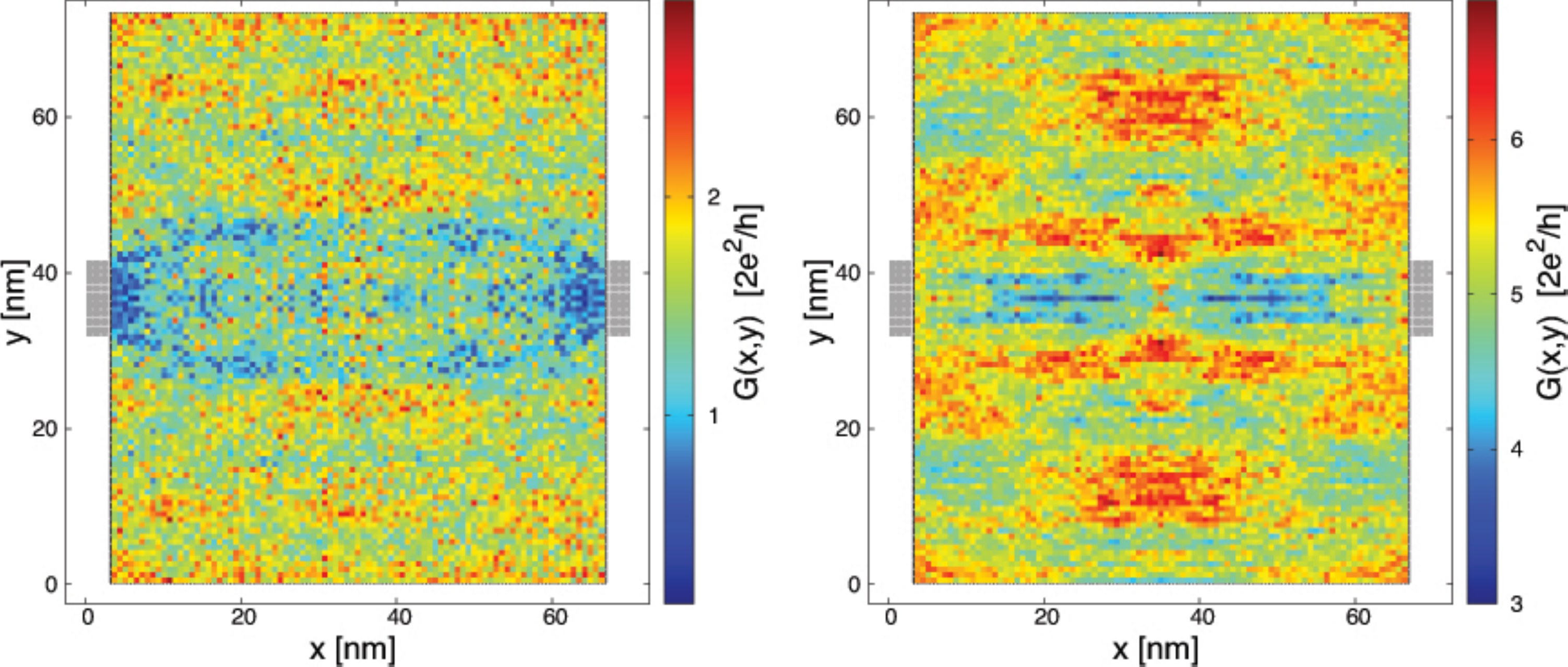}
\caption{\label{fig:FermiWavelength} (Color online)
Conductance maps for two Fermi energies, (left) $E_F=0.1 \gamma$ and (right) $E_F=0.9 \gamma$, corresponding to Fermi wavelengths of 13 nm and 1.5 nm, respectively. 
}
\end{center}	
\end{figure}

The Fermi wavelength in the system presented in Fig.~\ref{deltaY-scan} is $\lambda_F = 2.7$ nm.
Figure~\ref{fig:FermiWavelength} presents the simulated conductance maps for the same configuration of the leads for $E_F=0.1~\gamma$ and $E_F=0.9~\gamma$, corresponding to Fermi wavelengths of 13 nm and 1.5 nm.
The resolution resolved in the images, i.e., the size of the pixels in each of the maps, is also 1.5 nm. 
 
\begin{figure*}[t]
\begin{center}
\includegraphics[width=0.85\textwidth]{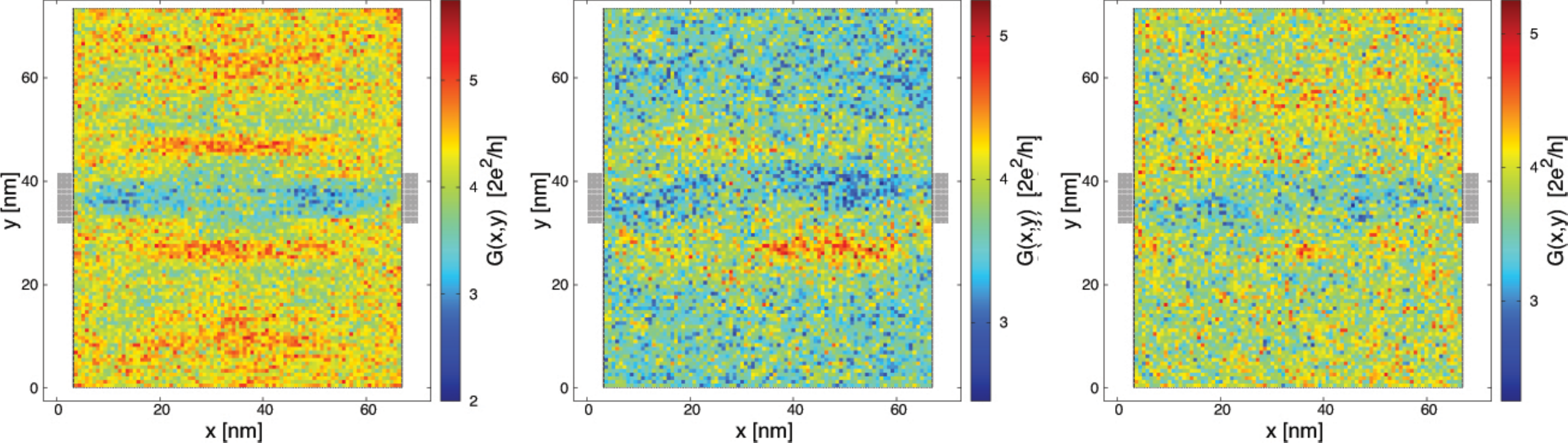}
\caption{\label{Clean_Edge_Bulk} (Color online)
SPM conductance maps simulations for disordered systems. 
(Left) Graphene sample with disordered edges and (middle, right) devices with bulk disorder.
While the interference pattern survives in devices with edge disorder, it is not present in systems with bulk disorder. 
At a mean free path of $l = 108$ nm, larger than the system size, $L=65$ nm, the mobility of the disordered devices presented in the middle panel is comparable to that of devices in crystalline substrates. 
The simulation presented in (c) is for a system with $l = 24$ nm. 
Other parameters are as in Fig.~\ref{deltaY-scan}.}
\end{center}	
\end{figure*}

Previous analytical results of Braun {\em et al.}~\cite{Braun_2008} consider ballistic trajectories between two constrictions in the presence of the tip scattering potential.
To first order, as treated in Ref.~\onlinecite{Braun_2008}, there are only two trajectories that interfere with each other: the source-to-drain trajectory and the source-tip potential-drain trajectory, revealing an interference pattern in the conductance maps.
In particular, the first order calculation predicts that the interference pattern depends on the Fermi wavelength of the system.
Comparing the conductance maps for three different energies (Fig.~\ref{deltaY-scan}(d) and those in Fig.~\ref{fig:FermiWavelength}), the interference pattern are very similar.
Increasing the Fermi energy of the system does not reveal the Fermi wavelength of the particles but rather results in better resolution of the interference patterns.
In contrast to analytical results,~\cite{Braun_2008} our numerical calculation treats all possible trajectories including reflections from the edge of the device and multiple scattering events.
We believe that the pattern found is due to the redirection of the charge carriers as the tip scans the device and does not dependent on the wavelength but rather on the geometry of the system and the interference of several trajectories the charge carriers can take as they traverse the device. 

In what follows we study the effect of edge and bulk disorder on the simulated conductance maps.
In Fig.~\ref{Clean_Edge_Bulk}, we present conductance maps for graphene devices with edge disorder (left panel) and bulk disorder (middle and right panels).
As seen in the left panel, the interference pattern is present in edge-disordered graphene devices. 
The effect of disorder at the edges of the device induces random scattering, modifying previously interfering trajectories and thus reducing the strength of the pattern.
This is more noticeable near the edge of the device.
Still, the result for edge-disordered systems is a comparable pattern found for pristine graphene (Fig.~\ref{deltaY-scan}(d)).

Our calculations show an important difference for bulk-disordered systems.
Surprisingly, the SPM simulations indicate diffusive rather than ballistic transport (in the mesoscopic systems considered) even on the regime where the mean free path (estimated from the Born approximation) is longer than the system size. 
This result does not question the validity of the Born approximation; rather, it shows the sensitivity of the SPM method for the geometry studied. 
Even when the size of the system is smaller than $l$, for this geometry, the oath traveled by the charge carrier from initially entering the device to exiting at the drain lead can be significantly longer. 
Our calculations take into account all possible trajectories that the charge carriers can take, including the possibility of several reflections from the leads and multiple scattering events from the walls of the device.
The presence of random disorder due to charged impurities in the substrate results in conductance maps with no interference patterns, (Fig.~\ref{Clean_Edge_Bulk}, middle and right panels).
The middle panel of Fig.~\ref{Clean_Edge_Bulk} is the conductance map for a system where the mean free path ($l = 107.5$ nm) is larger than the size of the device ($L = 65$ nm).
This value of $l$ is attainable in graphene devices with a crystalline BN-substrate, resulting in relatively high mobility~\cite{Dean_2010}. 
The mean free path of the system pictured in the right panel is $l = 24$ nm, comparable to high quality graphene devices in a silicon oxide substrate. 

Ballistic transport is expected when the mean free path is larger than the size of the system.
In contrast, in graphene devices with weak bulk disorder such that $l > L$, we find diffusive transport signatures.
This indicates that for the system and sizes considered, multiple reflections of the Dirac electrons from the edges dominate.
This mesoscopic effect needs to be carefully considered in both theoretical and experimental studies on mesoscopic graphene flakes.
As a result, SPM imaging could be used to explore the ballistic to diffusive transition in graphene devices.
The visible interference pattern in the conductance maps would be robust against the presence of edge disorder, inevitable in most experimental realizations, but is sensitive to the presence of disorder due to bulk disorder in the system.
Finally, we would like to comment on possible experimental realizations of SPM measurements in graphene.
Several SPM measurements have been carried out for graphene devices on a SiO$_2$ substrate~\cite{Berezovsky:2010wl,Berezovsky:2010wr,Schnez_2010} and can be readily carried out in suspended graphene, where the ballistic transport regime can be reached, in graphene membranes,~\cite{Wang_2012} and in devices with crystalline substrate, where the ballistic to diffusive cross-over should be observable.

\section{Conclusion}

We have proposed a method to study the mechanisms of electronic transport based on displacement conductance, that is, the conductance as a function of the position of the drain lead in a two-terminal device. 
The method extracts the conductance profile of the charge flow at the edge of the device and it can be applied to discern the mechanism of transport from ballistic to diffusive.
It is worth noting that this technique might be capable of uncovering signatures of anomalous transport in mesoscopic systems.  
Several quantum interference effects (Fabry-P\'erot resonances and UCFs) are also found in the resulting conductance curves.
The method that we applied to graphene is general, resolves the transport mechanism in graphene which cannot be accomplished via Fano shot noise, and can be used as well to study other systems such as semiconductor heterostructures. 

In this article, we simulated SPM measurements for the same devices considered in the conductance profile study.
Probe imaging is important as we are able to visualize the flow of charge and do not have to rely on transport metrics.
For the graphene devices considered, the SPM simulations suggest diffusive rather than ballistic transport even in the ``clean" regime where the mean free path is larger than the systems size.
Our numerical results suggest that in the process of escaping the device, charge carriers perform multiple reflections.
From our simulations, we expect that SPM measurements are well suited to study the crossover between ballistic and diffusive transport in graphene devices.

\section{acknowledgments}

We are grateful to S.~Barraza-Lopez, S.~Bhandari, J. Berezovsky, and R.~M.~Westervelt for valuable discussions. 
M.F.B. and E.J.H. were supported by the Department of Energy, Office of Basic Science (Grant No.~DE-FG02-08ER46513), and H.H. acknowledges funding through the German Research Foundation (Grant No.~HE 6312/1-1).
While most of the calculations were run on the Odyssey cluster supported by the FAS at Harvard University, the rest were performed on the Vaquero cluster supported by Oklahoma State University and outfitted with a Tesla K20 GPU donated by the NVIDIA Corporation.

\bibliographystyle{unsrt}

\begin{thebibliography}{24}
\expandafter\ifx\csname natexlab\endcsname\relax\def\natexlab#1{#1}\fi
\expandafter\ifx\csname bibnamefont\endcsname\relax
 \def\bibnamefont#1{#1}\fi
\expandafter\ifx\csname bibfnamefont\endcsname\relax
 \def\bibfnamefont#1{#1}\fi
\expandafter\ifx\csname citenamefont\endcsname\relax
 \def\citenamefont#1{#1}\fi
\expandafter\ifx\csname url\endcsname\relax
 \def\url#1{\texttt{#1}}\fi
\expandafter\ifx\csname urlprefix\endcsname\relax\def\urlprefix{URL }\fi
\providecommand{\bibinfo}[2]{#2}
\providecommand{\eprint}[2][]{\url{#2}}

\bibitem[{\citenamefont{Wallace}(1947)}]{Wallace_1947}
\bibinfo{author}{\bibfnamefont{P.~R.} \bibnamefont{Wallace}},
 \bibinfo{journal}{Phys. Rev.} \textbf{\bibinfo{volume}{71}},
 \bibinfo{pages}{622} (\bibinfo{year}{1947}).

\bibitem[{\citenamefont{CastroNeto_2009}(2009)}]{CastroNeto_2009}
A.~H.~Castro Neto, F.~Guinea, N.~M.~R.~Peres, K.~S.~Novoselov, and A.~K.~Geim, 
Rev. Mod. Phys. {\bf 81}, 109 (2009).

\bibitem[{\citenamefont{DasSarma_2011}(2011)}]{DasSarma_2011}
S.~Das Sarma, S.~ Adam, E.~H.~Hwang, and E.~Rossi, 
Rev. Mod. Phys. {\bf 83}, 407 (2011).

\bibitem[{\citenamefont{Geim_2009}(2009)}]{Geim_2009}
A.~K. Geim,  
Science {\bf 324}, 1530 (2009).

\bibitem[{\citenamefont{Rozhkov_2011}(2011)}]{Rozhkov_2011}
A.~V.~Rozhkov, G.~Giavaras, Y.~P.~Bliokh, V.~Freilikher, and F.~Nori,
Phys. Rep. {\bf 503}, 77 (2011).

\bibitem[{\citenamefont{Novoselov_2004}(2004)}]{Novoselov_2004}
K.~S.~Novoselov, A.~K. Geim, S.~V. Morozov, D. Jiang, Y. Zhang, S.~V. Dubonos, I.~V. Grigorieva, and A.~A. Firsov, 
Science {\bf 306}, 666 (2004).

\bibitem[{\citenamefont{Berger_2006}(2006)}]{Berger_2006}
C.~Berger, {\em et al.}, 
Science {\bf 312}, 1191 (2006).

\bibitem[{\citenamefont{Meyer_2007}(2007)}]{Meyer_2007}
J.~C.~Meyer, A.~K. Geim, M.~I.~Katsnelson, K.~S.~Novoselov, T.~J.~Booth, and S.~Roth, 
Nature {\bf 446}, 60 (2007).

\bibitem[{\citenamefont{Bolotin_2008}(2008)}]{Bolotin_2008}
K.~I.~Bolotin, K.~J.~Sikes, Z. Jiang, G. Fudenberg, J. Hone, P. Kim, and H.~L.~Stormer,
Solid State Commun. {\bf 146}, 351 (2008)

\bibitem[{\citenamefont{Du_2008}(2008)}]{Du_2008}
X.~Du, I.~Skachko, A.~Barker, and E.~Y.~Andrei, 
Nature Nano. {\bf 3}, 491 (2008).

\bibitem[{\citenamefont{Dean_2010}(2010)}]{Dean_2010}
C.~R.~Dean, {\em et al.}, 
Nature Nano. {\bf 5}, 722 (2010).

\bibitem[{\citenamefont{Adam_2007}(2007)}]{Adam_2007}
S.~ Adam, E.~H.~Hwang, V.~Galitski, and S.~Das Sarma, 
Proc. Natl. Acad. Sci. {\bf 104}, 18392 (2007).

\bibitem[{\citenamefont{Katsnelson_2007}(2007)}]{Katsnelson_2007}
M.~I.~Katsnelson and A.~K. Geim, 
Phil. Trans. R. Soc. A {\bf 366}, 195 (2007).

\bibitem[{\citenamefont{Gibertini_2010}(2010)}]{Gibertini_2010}
M. Gibertini, A. Tomadin, M. Polini, A. Fasolino, and M.~I.~Katsnelson,  
Phys. Rev. B {\bf 81}, 125437 (2010).

\bibitem[{\citenamefont{Tworzydlo_2006}(2006)}]{Tworzydlo_2006}
J.~Tworzydlo, B.~Trauzettel, M.~Titov, A.~Rycerz, and C.~W.~J.~Beenakker, 
Phys. Rev. Lett. {\bf 96}, 246802 (2006).

\bibitem[{\citenamefont{Miao_2007}(2007)}]{Miao_2007}
F.~Miao, S.~Wijeratne, Y.~Zhang, U.~C.~Coskun, W.~Bao, and C.~N.~Lau, 
Science {\bf 317}, 1530 (2007).

\bibitem[{\citenamefont{DiCarlo_2008}(2008)}]{DiCarlo_2008}
L.~DiCarlo, J.~R.~Williams, Y.~Zhang, D.~T.~McClure, and C.~M.~Marcus, 
Phys. Rev. Lett. {\bf 100}, 156801 (2008).

\bibitem[{\citenamefont{Danneau_2008}(2008)}]{Danneau_2008}
R.~Danneau, F.~Wu, M.~F.~Craciun, S.~Russo, M.~Y.~Tomi, J.~Salmilehto, A.~F.~Morpurgo, and P.~J.~Hakonen, 
Phys. Rev. Lett. {\bf 100}, 196802 (2008).

\bibitem[{\citenamefont{Blanter_2000}(2000)}]{Blanter_2000}
Y.~M.~Blanter and M.~B\"uttiker, 
Phys. Rep. {\bf 336}, 1 (2000).

\bibitem[{\citenamefont{Beenakker_1992}(1992)}]{Beenakker_1992}
C.~W.~J.~Beenakker and M.~B\"uttiker, 
Phys. Rev. B {\bf 46}, 1889 (1992).

\bibitem[{\citenamefont{Schomerus_2007}(2007)}]{Schomerus_2007}
H.~Schomerus, 
Phys. Rev. B {\bf 76}, 045433 (2007).

\bibitem[{\citenamefont{Barraza_2012}(2012)}]{Barraza_2012}
S.~Barraza-Lopez, M.~Kindermann, and M.~Y.~Chou, 
Nano Lett. Article ASAP, (2012).

\bibitem[{\citenamefont{Louis_2007}(2007)}]{Louis_2007}
E.~Louis, J.~A.~Verges, F.~Guinea, and G.~Chiappe, 
Phys. Rev. B {\bf 75}, 085440 (2007).

\bibitem[{\citenamefont{San-Jose_2007}(2007)}]{San-Jose_2007}
P.~San-Jose, E.~Prada, and D.~S.~Golubev, 
Phys. Rev. B {\bf 76}, 195445 (2007).

\bibitem[{\citenamefont{Lewenkopf_2008}(2008)}]{Lewenkopf_2008}
C.~H.~Lewenkopf, E.~R.~Mucciolo, and A.~H. CastroNeto, 
Phys. Rev. B {\bf 77}, 081410(R) (2008).

\bibitem[{\citenamefont{Cheianov_2006}(2006)}]{Cheianov_2006}
V.~V.~Cheianov and V.~I.~Fal'ko, 
Phys. Rev. B {\bf 74}, 041403(R) (2006).

\bibitem[{\citenamefont{Topinka_2000}(2000)}]{Topinka_2000}
M.~A.~Topinka, B.~J.~LeRoy, S.~E.~J.~Shaw, E.~J.~Heller, R.~M.~Westervelt, K.~D.~Maranowski, and A.~C.~Gossard, 
Science {\bf 289}, 2323 (2000).

\bibitem[{\citenamefont{Topinka_2001}(2001)}]{Topinka_2001}
M.~A.~Topinka, B.~J.~LeRoy, R.~M.~Westervelt, S.~E.~J.~Shaw, R. Fleischmann, E.~J.~Heller,  K.~D.~Maranowski, and A.~C.~Gossard, 
Nature {\bf 410}, 183 (2001).

\bibitem[{\citenamefont{Barthelemy_2008}(2008)}]{Barthelemy_2008}
P.~Barthelemy, J.~Bertolotti, and D.~S.~Wiersma,
Nature {\bf 453}, 495 (2008).

\bibitem[{\citenamefont{Haug_2008}(2008)}]{Haug_2008}
H.~Haug and A.-P.~Jaujo, 
{\em Quantum Kinetics in Transport and Optics of Semiconductors}, 
2nd Ed. (Springer, Berlin, 2008).

\bibitem[{\citenamefont{Robinson_2007}(2007)}]{Robinson_2007}
J.~P.~Robinson and H.~Schomerus, 
Phys. Rev. B {\bf 76}, 115430 (2007).

\bibitem{Rycerz_2007}
\bibinfo{author}{\bibfnamefont{A.} \bibnamefont{Rycerz}},
\bibinfo{author}{\bibfnamefont{J.} \bibnamefont{Tworzydlo}},
 \bibnamefont{and}
 \bibinfo{author}{\bibfnamefont{C.~W.~J.} \bibnamefont{Beenakker}},
  \bibinfo{journal}{EPL} \textbf{\bibinfo{volume}{79}},
  \bibinfo{pages}{57003} (\bibinfo{year}{2007}).

\bibitem[{\citenamefont{Suzuura_2002}(2002)}]{Suzuura_2002}
H.~Suzuura and T.~Ando, 
Phys. Rev. Lett. {\bf 89}, 266603 (2002).

\bibitem{Nicolic_2010}
B.~K.~Nicolic, L.~P.~Zarbo, and S.~Souma, 
Spin currents in Semiconductor nanostructures: a nonequilibrium Green-function approach,
In A.~V.~Narlikar and Y.~Y.~Fu (Eds.), 
{\em The Oxford Handbook on Nanoscience and Technology Frontiers and Advances}, 
(Oxford University Press, Oxford, 2010).

\bibitem{Landauer_1957}
R. Landauer, IBM J. Res. Develop. 1, 233 (1957).

\bibitem[{\citenamefont{Huang_2009}(2009)}]{Huang_2009}
L.~Huang, Y.-C.~Lai, D.~K.~Ferry, S.~M.~Goodnick, and R. Akis, 
Phys. Rev. Lett., {\bf 103}, 054101 (2009).

\bibitem[{\citenamefont{Huang_2011}(2011)}]{Huang_2011}
L.~Huang, R.~Yang, and Y.-C.~Lai,
EPL, {\bf 94}, 58003 (2011).

\bibitem[{\citenamefont{Martin_2008}(2008)}]{Martin_2008}
J.~Martin, N.~Akerman, G.~Ulbricht, T.~Lohmann, J.~H.~Smet, K. Von Klitzing, and A.~Yacoby, 
Nature. Phys. {\bf 4}, 144 (2008).

\bibitem[{\citenamefont{Lee_1985}(1985)}]{Lee_1985}
P.~A.~Lee and A.~D. Stone, 
Phys. Rev. Lett. {\bf 55}, 1622 (1985).

\bibitem[{\citenamefont{Altshuler_1985}(1985)}]{Altshuler_1985}
B.~L.~Altshuler and D.~E.~Khmelnitskii, 
JETP Lett. {\bf 42}, 359 (1985).

\bibitem[{\citenamefont{Feng_1986}(1986)}]{Feng_1986}
S.~Feng, P.~A.~Lee and A.~D.~Stone, 
Phys. Rev. Lett. {\bf 56}, 1960 (1986).

\bibitem[{\citenamefont{Altshuler_1985b}(1985)}]{Altshuler_1985b}
B.~L.~Altshuler and B.~Z.~Spivak, 
JETP Lett. {\bf 42}, 447 (1985).

\bibitem[{\citenamefont{Borunda_2011}(2011)}]{Borunda_2011}
\bibinfo{author}{\bibfnamefont{M.~F.} \bibnamefont{Borunda}},
 \bibinfo{author}{\bibfnamefont{J.}~\bibnamefont{Berezovsky}},
 \bibinfo{author}{\bibfnamefont{R.~M.} \bibnamefont{Westervelt}}, and
 \bibinfo{author}{\bibfnamefont{E.~J.} \bibnamefont{Heller}},
ACS Nano {\bf 5}, 3622 (2011).


\bibitem[{\citenamefont{Marcus et~al.}(1992)\citenamefont{Marcus, Rimberg,
  Westervelt, and Hopkins}}]{Marcus:1992ug}
\bibinfo{author}{\bibfnamefont{C.~M.}~\bibnamefont{Marcus}},
  \bibinfo{author}{\bibfnamefont{A.~J.}~\bibnamefont{Rimberg}},
  \bibinfo{author}{\bibfnamefont{R.~M.}~\bibnamefont{Westervelt}},
  \bibinfo{author}{\bibfnamefont{P.~F.}~\bibnamefont{Hopkins}},   \bibnamefont{and} \bibinfo{author}{\bibfnamefont{A.~C.}~\bibnamefont{Gossard}},
  \bibinfo{journal}{Phys. Rev. Lett.} \textbf{\bibinfo{volume}{69}},
  \bibinfo{pages}{506} (\bibinfo{year}{1992}).

\bibitem[{\citenamefont{Jalabert et~al.}(1990)\citenamefont{Jalabert, Baranger,
  and Stone}}]{Jalabert_90_PRL}
\bibinfo{author}{\bibfnamefont{R.~A.} \bibnamefont{Jalabert}},
  \bibinfo{author}{\bibfnamefont{H.~U.} \bibnamefont{Baranger}},
  \bibnamefont{and} \bibinfo{author}{\bibfnamefont{A.~D.} \bibnamefont{Stone}},
  \bibinfo{journal}{Phys. Rev. Lett.} \textbf{\bibinfo{volume}{65}},
  \bibinfo{pages}{2442} (\bibinfo{year}{1990}).

\bibitem[{\citenamefont{Baranger et~al.}(1993)\citenamefont{Baranger, Jalabert,
  and Stone}}]{Baranger:1993wl}
\bibinfo{author}{\bibfnamefont{H.}~\bibnamefont{Baranger}},
  \bibinfo{author}{\bibfnamefont{R.}~\bibnamefont{Jalabert}}, \bibnamefont{and}
  \bibinfo{author}{\bibfnamefont{A.}~\bibnamefont{Stone}},
  \bibinfo{journal}{Chaos} \textbf{\bibinfo{volume}{3}}, \bibinfo{pages}{665}
  (\bibinfo{year}{1993}).

\bibitem[{\citenamefont{Ponomarenko et~al.}(2008)\citenamefont{Ponomarenko,
  Schedin, Katsnelson, Yang, Hill, Novoselov, and Geim}}]{Ponomarenko:2008ty}
\bibinfo{author}{\bibfnamefont{L.}~\bibnamefont{Ponomarenko}},
  \bibinfo{author}{\bibfnamefont{F.}~\bibnamefont{Schedin}},
  \bibinfo{author}{\bibfnamefont{M.}~\bibnamefont{Katsnelson}},
  \bibinfo{author}{\bibfnamefont{R.}~\bibnamefont{Yang}},
  \bibinfo{author}{\bibfnamefont{E.}~\bibnamefont{Hill}},
  \bibinfo{author}{\bibfnamefont{K.}~\bibnamefont{Novoselov}},
  \bibnamefont{and} \bibinfo{author}{\bibfnamefont{A.}~\bibnamefont{Geim}},
  \bibinfo{journal}{Science} \textbf{\bibinfo{volume}{320}},
  \bibinfo{pages}{356} (\bibinfo{year}{2008}).

\bibitem[{\citenamefont{Berezovsky et~al.}(2010)\citenamefont{Berezovsky,
  Borunda, Heller, and Westervelt}}]{Berezovsky:2010wl}
\bibinfo{author}{\bibfnamefont{J.}~\bibnamefont{Berezovsky}},
  \bibinfo{author}{\bibfnamefont{M.~F.} \bibnamefont{Borunda}},
  \bibinfo{author}{\bibfnamefont{E.~J.} \bibnamefont{Heller}},
  \bibnamefont{and} \bibinfo{author}{\bibfnamefont{R.~M.}
  \bibnamefont{Westervelt}}, \bibinfo{journal}{Nanotechnology}
  \textbf{\bibinfo{volume}{21}}, \bibinfo{pages}{274013}
  (\bibinfo{year}{2010}).

\bibitem[{\citenamefont{Berezovsky and Westervelt}(2010)}]{Berezovsky:2010wr}
\bibinfo{author}{\bibfnamefont{J.}~\bibnamefont{Berezovsky}} \bibnamefont{and}
  \bibinfo{author}{\bibfnamefont{R.~M.} \bibnamefont{Westervelt}},
  \bibinfo{journal}{Nanotechnology} \textbf{\bibinfo{volume}{21}},
  \bibinfo{pages}{274014} (\bibinfo{year}{2010}).

\bibitem[{\citenamefont{Schnez_2010}(2010)}]{Schnez_2010}
S.~Schnez, J.~G\"uttinger, M.~Huefner, C.~Stampfer, K.~Ensslin, and T.~Ihn, 
Phys. Rev. B {\bf 82}, 165445 (2010).

\bibitem[{\citenamefont{Braun_2008}(2008)}]{Braun_2008}
M.~Braun, L.~Chirolli, and G.~Burkard,
Phys. Rev. B {\bf 77}, 115433 (2008).


\bibitem[{\citenamefont{Wang_2012}(2012)}]{Wang_2012}
W.~L.~Wang, S.~Bhandari, W.~Yi, D.~C.~Bell, R.~Westervelt, and E.~Kaxiras, 
Nano Lett. {\bf 12}, 2278 (2012).

\end{thebibliography}

\end{document}